\documentclass[titlepage,12pt]{article}
\usepackage{amssymb,psfig,epsfig,pslatex}
\def\sp{{\mathbf s}_t}
\newcommand{\sm}{{\mathbf s}_{\bar{t}}}
\newcommand{\kh}{{\hat{\mathbf k}}}
\newcommand{\ph}{\hat{\mathbf p}}
\newcommand{\dhh}{\hat{\mathbf d}}
\newcommand{\one}{1\!\mbox{l}}
\def\percent{\%}
\def\as{{\alpha_{s}}}
\def\Eq#1{Eq.~(\ref{#1})}
\def\peters#1{{ \it \underline{}}}

\begin{document}
\begin{titlepage}
\noindent DESY-01-085 \hfill July 2001\\
PITHA 01/06 \hfill \\
TTP01-16 \hfill \\
\vspace{0.01cm}
\renewcommand{\thefootnote}{\fnsymbol{footnote}}
\begin{center}
{\LARGE {\bf Top Quark Spin
Correlations
at Hadron \\ Colliders: Predictions at Next-to-Leading \\
\vskip 0.2cm Order QCD}\footnote{supported by BMBF, contract 05 HT1 PAA 4}} \\
\vspace{2cm}
{\bf W. Bernreuther\,$^{a}$,
A. Brandenburg\,$^{b,}$\footnote{supported by a Heisenberg fellowship of
D.F.G.},
Z. G. Si\,$^{a}$
and P. Uwer\,$^c$}
\par\vspace{1cm}
$^a$ Institut f\"ur\ Theoretische Physik, RWTH Aachen, 52056 Aachen, Germany\\
$^b$  DESY-Theorie, 22603 Hamburg, Germany\\
$^c$ Institut f\"ur Theoretische Teilchenphysik, Universit\"at Karlsruhe,
76128 Karlsruhe, Germany
\par\vspace{3cm}
{\bf Abstract:}\\
\parbox[t]{\textwidth}
{The collider experiments at the Tevatron and the LHC will
allow for  detailed investigations of the  properties of the  top quark.
This requires precise predictions of
the hadronic production of $t\bar t$
pairs and of their subsequent decays.
In this Letter we present for the reactions
$p {\bar p}, p p \rightarrow t{\bar t}  + X \rightarrow \ell^+ \ell'^- +
X$ the first calculation of the dilepton angular distribution at
next-to-leading
order (NLO) in the QCD coupling,  keeping the full
dependence on the spins of the intermediate $t\bar{t}$ state.  
The angular distribution reflects the degree of  correlation of the
$t$ and $\bar t$ spins which
we determine  for different choices of $t$ and $\bar t$ spin bases.
In the case of the Tevatron,
the QCD corrections 
are sizeable, and the distribution 
is quite sensitive to the parton content of the proton.}
\end{center}
\vspace{1cm}
\end{titlepage}

\renewcommand{\thefootnote}{\arabic{footnote}}

\setcounter{footnote}{0}
The top quark is by far the heaviest fundamental fermion 
discovered \cite{tdisc} to date. It is an excellent probe
of the fundamental interactions in the high energy regime that will be
explored by the
upgraded Fermilab Tevatron collider and by the CERN large
hadron collider LHC.
It is expected that very large numbers of top quarks will be
produced with these colliders: eventually 
about 10$^4$ top quark-antiquark ($t\bar t)$
pairs per year at the Tevatron and  more than about 10$^7$
$t\bar t$ pairs per year at the LHC.
This will make feasible precise investigations of the 
interactions of top quarks.
\par
Because of their extremely short lifetime 
top quarks find no time to form hadronic
bound states: they are highly
instable particles whose interactions
are governed by short-distance dynamics \cite{Bigi:1986jk}.
As a consequence the properties of the top quark and antiquark, 
in particular phenomena
associated with their spins,
are reflected directly
in the distributions and the corresponding angular correlations 
of the jets, $W$ bosons, or
leptons into which the $t$ and $\bar t$ decay.
These distributions are determined 
by the  $t$ and $\bar t$ polarizations and
spin correlations induced by the production mechanism(s). Furthermore 
they depend on the
interactions responsible for the top (anti-)quark decay.
Hence the analysis of these distributions
will be an important tool, once large data samples will be
available, to obtain detailed information about 
top-quark production and decay.
\par
For hadronic pair production the spin
correlations of $t\bar t$ pairs were studied to leading order in the
coupling $\as$ of Quantum Chromodynamics (QCD) in 
ref.~\cite{study,Mahlon:1997}.
In particular it was analyzed which spin bases are most suitable for
the investigation of $t\bar t$ spin correlations induced by the strong
interactions{\footnote{A first attempt to measure $t\bar t$ spin correlations
with a very small data sample was made in ref.~\cite{Abbott:2000dt}.}}. 
There exists also an extensive literature,
for example \cite{newphys} and references therein,
on how to exploit top-quark spin phenomena at hadron colliders in the
search for new interactions.
The work  which we report in this Letter serves the purpose of putting
predictions
of $t\bar t$ spin correlations within the standard model of particle
physics (SM) on firmer grounds.
We analyze the hadronic production of $t\bar t$ pairs
and their subsequent decays, keeping the full information
on the spin configuration of the $t\bar t$ state.  
We extend the existing results by taking into account the next-to-leading 
order (NLO) QCD corrections in the production and the decay of the $t\bar t$
pairs.
More specifically we consider
the channels where  both $t$ and $\bar t$ decay semileptonically,
\begin{equation}
p {\bar p}, p p \rightarrow t{\bar t}  + X \rightarrow \ell^+ \ell\,'^- + X,
\label{eq:ttsemi}
\end{equation}
$(\ell = e,\mu,\tau),$ and we predict 
the following double leptonic distribution
at NLO in the coupling
$\alpha_s$:
\begin{eqnarray}
\frac{1}{\sigma}\frac{d^2\sigma}{d\cos\theta_+ d\cos\theta_-}=
\frac{1}{4} (1 +
{\rm B}_1\cos\theta_+
+ {\rm B}_2\cos\theta_- 
-{\rm C}\cos\theta_+ \cos\theta_-)\,\, ,
\label{eq:ddist1}
\end{eqnarray}
with $\sigma$ being the cross section for the channel under consideration.
In \Eq{eq:ddist1}  $\theta_+$ ($\theta_-$) denotes the angle between the
direction of flight of the lepton $\ell^+$ ($\ell\,'^-$) in the $t$ ($\bar{t}$)
rest frame{\footnote{
We define the rest frame of the $t (\bar t)$
quark by a rotation-free Lorentz boost from the  center-of-mass (c.m.)
frame of the initial partons that produce the $t\bar t$ pair.
If one defines the $t (\bar t)$ rest frame by a boost
from the hadronic c.m. frame, it will differ from our choice by a Wigner
rotation.}} and a reference 
direction $\hat{\bf a}$ ($\hat{\bf b}$). The directions
$\hat{\bf a}$, $\hat{\bf b}$ can be 
chosen arbitrarily. Different choices will yield 
different values for the coefficients 
${\rm B}_{1,2}$ and C.
The physical interpretation of these coefficients is well known  
\cite{study,Mahlon:1997}:
The coefficient C in \Eq{eq:ddist1}
reflects spin correlations of the $t\bar t$ intermediate
state. A more detailed discussion will be given below \Eq{eq:ddist2}.
For our choices of
the directions $\hat{\bf a}$ and $\hat{\bf b}$ (cf. \Eq{eq:spbasis})
QCD interactions yield vanishing  
coefficients ${\rm B}_1$, ${\rm B}_2${\footnote{
This is due to the parity invariance of QCD.}}. 
\par
In principle
one could measure the angular distribution of every possible 
decay product of the top (anti-)quark. 
In the SM, where the main top-quark decay
modes are $t\to b W \to b q {\bar q}',  b \ell \nu_{\ell}$, 
the most powerful analyzers of
the polarization of the top quark are the charged leptons, 
or the jets that originate from quarks of weak isospin
$-1/2$ produced by the decay of the $W$ boson. Here we restrict
ourselves to the double leptonic distribution.

To predict the ``dilepton + jets'' distribution (\ref{eq:ddist1}) 
at NLO accuracy 
we have to consider the following
parton subprocesses:
\begin{equation}
gg, q{\bar q} \ {\buildrel
 t{\bar t}\over \longrightarrow} \  b {\bar b} \ell^+
 \ell'^- \nu_{\ell} {\bar \nu}_{\ell'},
\label{eq:ttrec1}
\end{equation}
\begin{equation}
gg, q{\bar q} \  {\buildrel
 t{\bar t}\over \longrightarrow} \  b {\bar b} \ell^+
 \ell'^- \nu_{\ell} {\bar \nu}_{\ell'} + g,
\label{eq:ttrec2}
\end{equation}
\begin{equation}
g + q ({\bar q})\  {\buildrel
 t{\bar t}\over \longrightarrow}\   b {\bar b} \ell^+
 \ell'^- \nu_{\ell} {\bar \nu}_{\ell'} + q ({\bar q}) .
\label{eq:ttrec3}
\end{equation}
At the Tevatron the cross section is dominated by quark-antiquark 
annihilation while at the LHC gluon-gluon fusion is predicted to be
the dominant production process.

In view of the fact that the total width
$\Gamma_t$
of the top quark is much smaller than its mass $m_t$ 
($\Gamma_t/m_t ={\cal O}(1\%)$), one may  analyze the above reactions
using the so-called  leading pole approximation
\cite{Stuart:1991}. This amounts to
expanding the amplitudes of Eqs.~(\ref{eq:ttrec1}) - (\ref{eq:ttrec3})
around the poles
of the unstable $t$ and
$\bar t$ quarks. Only the leading term of this
expansion, i.e.,
the residue of the double poles is  kept here.  The radiative
corrections to the respective lowest-order amplitudes can be classified
into so-called
factorizable and non-factorizable
corrections.  We take into account  the factorizable
 corrections to the above reactions for which
the squared
 matrix element ${\cal M}$ is
of the form
$|{\cal M}|^2 \propto {\rm Tr}[\rho
R{\bar{\rho}}] .$
Here $R$ denotes the respective spin density matrix for the production of
on-shell $t\bar t$ pairs, and $\rho$ ($\bar\rho$) 
is the $t$ ($\bar t$) decay density matrix.

To obtain a theoretical prediction for the distribution in
\Eq{eq:ddist1} at NLO
accuracy we use our recent results \cite{Bernreuther:2000yn}
on the $t\bar t$ production spin-density matrices at NLO QCD.
These results extend previous calculations \cite{Nason:1988} of 
the differential $t\bar t$ cross section with spins summed over 
and allow the calculation of the cross section for a specific spin
configuration of the $t\bar t$ state. 
In particular, the quantization axes can be chosen
arbitrarily.

The decay density matrix $\rho$ (${\bar{\rho}}$) required for 
computing  (\ref{eq:ddist1}) describes the normalized angular 
distribution of the decay of a polarized $t
(\bar t)$ quark into $\ell^+ (\ell^-) + anything$ in the  
rest frame of the $t(\bar t)$ quark.
The matrix $\rho$  has the form
$2\rho_{\alpha'\alpha}
= (\one +{\kappa}_+\,\mathbf{\sigma} \cdot {\hat{\bf{q}}_+})_{\alpha'\alpha}$
where $\hat{\bf{q}}_+$
describes the direction of flight of $\ell^+$ in the rest frame of the $t$
quark and $\sigma_i$ denote the Pauli matrices.  
The decay matrix $\bar{\rho}$ is obtained from $\rho$ 
by replacing  $\hat{\bf{q}}_+$ by $-\hat{\bf{q}}_-$ and
${\kappa}_+$ by ${\kappa}_-$.
The factor ${\kappa}_+$ (${\kappa}_-$) signifies the top-spin 
analyzing power of the charged lepton. It is
equal to one to lowest order in the SM, that is, for $V-A$ charged currents.
Its value including the order $\alpha_s$ corrections can be extracted from
the results of \cite{Czarnecki:1991} and turns out to be very close 
to one: ${\kappa}_+ = {\kappa}_- =1-0.015 \alpha_s$.
Using the general expressions for
$\rho$, $\bar{\rho}$ and the fact that the factorizable contributions
are of the form ${\rm Tr}[\rho R{\bar{\rho}}]$ one obtains the
following formula for the correlation coefficient $\rm C$ in 
\Eq{eq:ddist1}:
\begin{equation}
{\rm C} = 4 {\kappa}_+{\kappa}_- \langle (\hat{\bf{a}}\cdot
\sp)(\hat{\bf{b}}\cdot \sm) \rangle   ,
\label{eq:ddist2}
\end{equation}
where $\sp,\sm$ denote the $t$ and $\bar t$ spin operators.
The expectation value
in \Eq{eq:ddist2} is defined with respect to the matrix elements for
the hadronic production of $t\bar t X$.
It is related to the more familiar double spin asymmetries
\begin{equation}
4 \langle (\hat{\bf{a}}\cdot \sp)(\hat{\bf{b}}\cdot \sm) \rangle = {\rm
N(\uparrow \uparrow)+\rm N(\downarrow \downarrow)
  - \rm N(\uparrow \downarrow)- \rm N(\downarrow \uparrow)\over
  \rm N(\uparrow \uparrow)+\rm N(\downarrow \downarrow)
  + \rm N(\uparrow \downarrow)+ \rm N(\downarrow \uparrow)
  },
\label{eq:ddist3}
\end{equation}
where $N(\uparrow \uparrow)$ etc. denote the number
of $t\bar t$ pairs  with $t$ and $\bar t$ spin parallel -- or anti-parallel --
to  $\hat{\bf a}$ and $\hat{\bf b}$, respectively.
From \Eq{eq:ddist3} one can see that 
the axes $\hat{\bf a}$, 
$\hat{\bf b}$ introduced through the angles $\theta_\pm$ in \Eq{eq:ddist1}
can be interpreted as quantization axes of the intermediate $t\bar t$
state within our approximation.
\Eq{eq:ddist2} generalizes the lowest-order results of
\cite{study,Mahlon:1997}
and holds for factorizable contributions to all
orders in the QCD coupling{\footnote{The non-factorizable 
NLO QCD corrections were calculated for
$g g$ and $q \bar q$ initial states in ref.~\cite{Beenakker:1999}.
We expect  with these results that the effect of these corrections 
on the dileptonic angular correlations
is small.}}.
\par
For definiteness we consider here the following  spin bases:
\begin{equation}
  \begin{array}[h]{lll}
    \hat {\bf a} = \kh_t,&  \hat {\bf b} = \kh_{\bar t}&
    \mbox{(helicity basis)}, \\
    \hat {\bf a} = \ph,& \hat{\bf b} = \ph, &\mbox{(beam\
      basis)},\\
    \hat {\bf a} =\dhh_t,& \hat{\bf b} = \dhh_{\bar t}&
    \mbox{(off-diagonal\ basis)}.
  \end{array}
\label{eq:spbasis}
\end{equation}
Here $\kh_t  (\kh_{\bar t})$ denotes the direction of
flight of the $t (\bar t)$ quark in the parton c.m.s., and $\hat{\bf p}$ is 
the unit vector along one of the hadronic beams in the laboratory frame.
Furthermore $\dhh_t$ is the axis constructed
in ref.~\cite{Mahlon:1997} with respect to which the spins
of $t$ and $\bar t$ produced by $q{\bar q}$ annihilation are 100 \percent\
correlated{\footnote{
We use the  definitions for $\dhh_t$ and  $\dhh_{\bar t}$ given
in ref.~\cite{Bernreuther:2000yn}.
In particular,  $\dhh_{\bar t}=\dhh_t$ at LO.
The sign of $C_{\rm off.}$ at LO  is therefore
 opposite to that of \cite{Mahlon:1997}.
}} to leading order in $\alpha_s$.
(For $gg\to t\bar t$ one can show that no spin basis
with this property exists.)
\par
Table \ref{tab:cteq5}
contains our results{\footnote{We use the $\overline{\rm MS}$ 
factorization scheme, $\alpha_s$ is defined to be
the five-flavour $\overline{\rm MS}$ coupling, and $m_t$ is defined in the
on-shell scheme.} for ${\rm C}$ at leading and
next-to-leading order
in $\alpha_s$ using the parton distribution
functions (PDF) CTEQ5L (LO) and CTEQ5M (NLO) of \cite{CTEQ5}.
(These numbers and the results given
below were obtained by
integrating over the full phase phase. Results with cuts included
will be given elsewhere \cite{bbsu3}.) 
For $p \bar p$ collisions at $\sqrt{s}=$ 2 TeV
the helicity basis is not
the best choice  because the $t$, $\bar t$
quarks are only moderately relativistic in this case.
Table  \ref{tab:cteq5} shows that the dilepton spin correlations
at the Tevatron are large both in  the
off-diagonal and in the beam basis. In fact they are  almost identical.
The QCD corrections
decrease the LO  results for these correlations
by about 10\percent. Since the $gg$ initial state
dominates $t\bar t$ production
with  $p p$ collisions at 
$\sqrt{s}=$ 14 TeV  the beam and off-diagonal bases
are no longer useful.
Here the helicity
basis is a good choice and gives a spin correlation of about 30\percent.
In this case the QCD corrections are small.
The large difference between the LO and NLO results for the correlation
in the beam basis
at the LHC is due to an almost complete cancellation of the
contributions from the $q\bar{q}$ and $gg$ initial state at LO. 
\begin{table}[h]
\begin{center}\renewcommand{\arraystretch}{1.5}
\begin{tabular}{|c|cc|cc|} \hline
 &\multicolumn{2}{c|}{$p\bar p$ at $\sqrt{s}=2$~TeV }
&\multicolumn{2}{c|}{$pp$ at $\sqrt{s}=14$~TeV }
\\
\cline{2-5}
& LO & NLO & LO & NLO\\  \hline \hline
${\rm C}_{\rm hel.}$ & $-0.456$& $-0.389$ & $\hphantom{-}0.305$  & 
$\hphantom{-}0.311$\\
${\rm C}_{\rm beam}$ & $\hphantom{-}0.910$&  $\hphantom{-}0.806$ & 
 $-0.005$ & $-0.072$\\
${\rm C}_{\rm off.}$ & $\hphantom{-}0.918$ & $\hphantom{-}0.813$ & 
$-0.027$ & $-0.089$\\ \hline
\end{tabular}
\vspace*{1em}
\end{center}
\caption{\it Coefficient ${\rm C}$
of \Eq{eq:ddist2}
to leading (LO) and
next-to-leading order (NLO)  in $\alpha_s$ for the spin bases of 
\Eq{eq:spbasis}. The parton distribution
functions of \protect\cite{CTEQ5} were used choosing the renormalization scale
$\mu_R$ equal to the factorization scale $\mu_F$ =
$m_t$ = 175 GeV.} \label{tab:cteq5}
\end{table}
\begin{figure}[h]
\centerline{\unitlength 1cm
\begin{picture}(9,9)
\put(0,-1){\psfig{figure=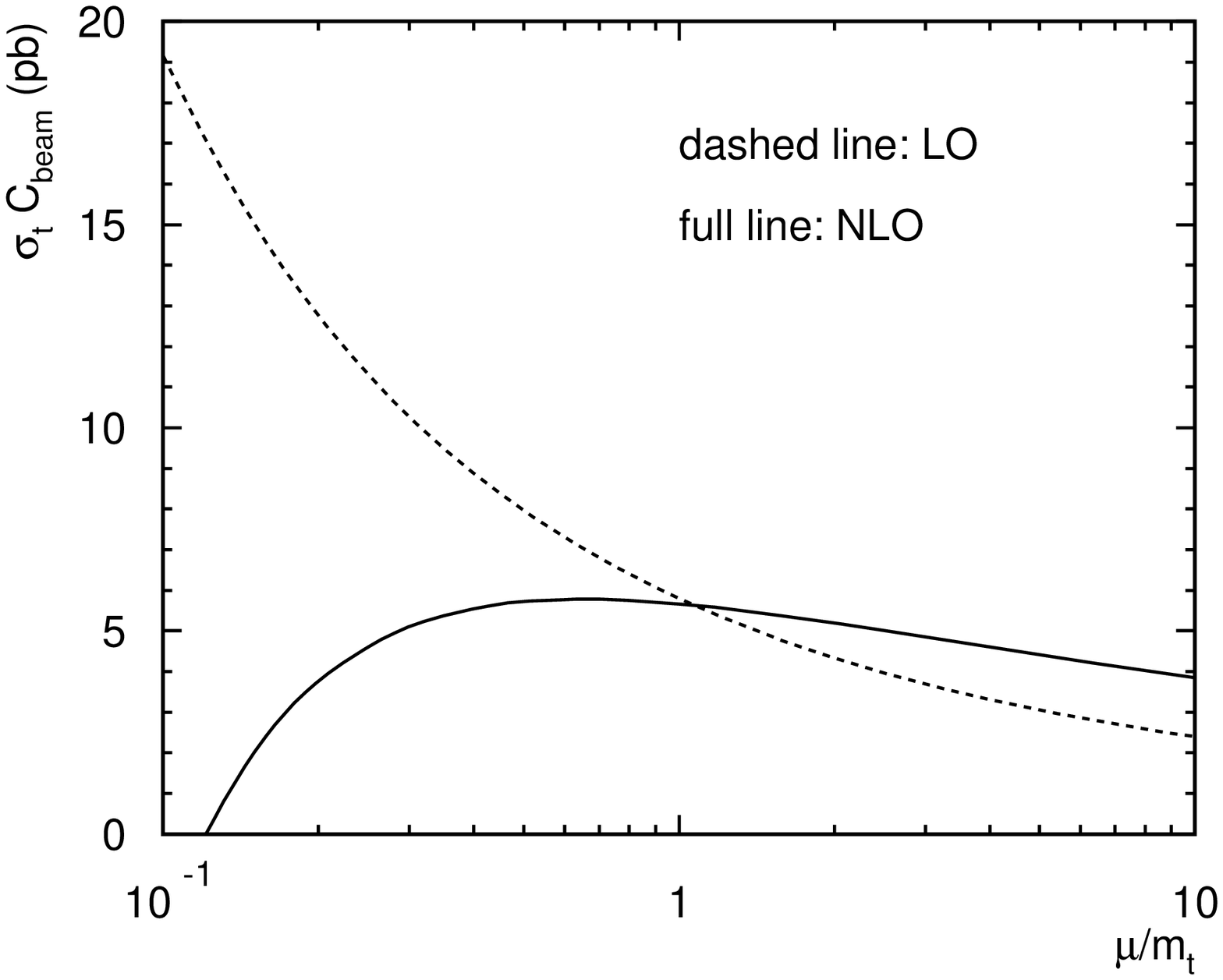,width=8cm,height=8cm}}
\end{picture}
}
\vskip -2cm
\caption{\it Dependence of
$\sigma_t{\rm C}_{\rm beam}$ at LO (dashed line) and  at NLO
(solid line) on $\mu=\mu_R=\mu_F$
for $p \bar p$ collisions at $\sqrt{s}=$ 2 TeV,
 with PDF  of  \protect\cite{CTEQ5}.}
\label{fig:mubeam}
\end{figure}

\begin{figure}[h]
\centerline{\unitlength 1cm
\begin{picture}(9,9)
\put(0,-2){\psfig{figure=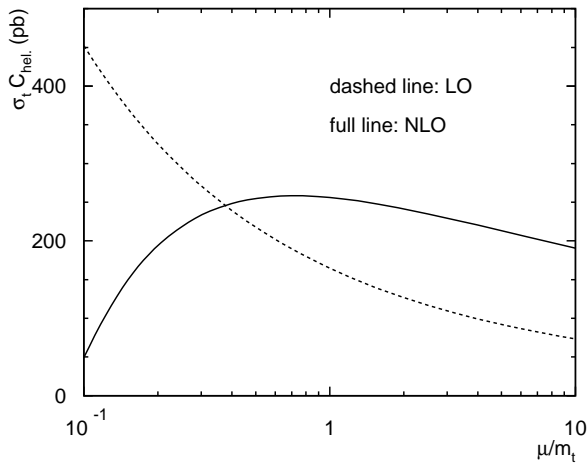,width=8cm,height=8cm}}
\end{picture}
}
\vskip -1cm
\caption{\it Dependence of
$\sigma_t{\rm C}_{\rm hel.}$ at LO (dashed line) and  at NLO
(solid line) on $\mu=\mu_R=\mu_F$
for $p  p$ collisions at $\sqrt{s}=$ 14 TeV,
with PDF  of  \protect\cite{CTEQ5}.}\label{fig:muhel}
\end{figure}

\par
We now discuss the uncertainties of our predictions.
It is well known that the inclusion of the QCD corrections
reduces the dependence of the $t\bar{t}$ cross section $\sigma_t$ on
the renormalization and factorization scales significantly.
The same is true
for  the product $\sigma_t {\rm C}$. 
In Figs.~\ref{fig:mubeam} and \ref{fig:muhel} 
we demonstrate this with
$\sigma_t {\rm C_{\rm beam}}$ and $\sigma_t {\rm C_{\rm hel.}}$ evaluated
at Tevatron and LHC energies, respectively,
as  functions of $\mu/m_t$, where $\mu=\mu_R=\mu_F$. The corresponding
figure for $\sigma_t {\rm C_{\rm off.}}$
is almost identical to Fig. \ref{fig:mubeam}.
\par
To leading order in $\alpha_s$ the coefficient C depends only
on the factorization scale $\mu_F$, while at NLO  it
depends on both scales $\mu_R$ and $\mu_F$. Table \ref{tab:mudep}
shows our NLO results for the three choices $\mu_R=\mu_F=m_t/2,m_t,2m_t$,
again using the PDF of \cite{CTEQ5}. An extension of this  work, which is
however beyond
the scope of this Letter,  would be the
resummation of Sudakov-type logarithms at the next-to-leading logarithmic
level. This was performed  in ref.~\cite{Kidonakis:1997gm} for the
total cross section  $\sigma_t$
and it stabilizes  the predictions for $\sigma_t$ with
respect to variations of $\mu_R$ and $\mu_F$.

\begin{table}[h]
\begin{center}\renewcommand{\arraystretch}{1.5}
\begin{tabular}{|c|ccc|c|} \hline
& \multicolumn{3}{c|}{$p\bar p$ at $\sqrt{s}=2$~TeV}
& $p p$ at $\sqrt{s}=14$~TeV \\
\cline{2-5}
 $\mu_R=\mu_F$   & ${\rm C}_{\rm hel.}$ &  ${\rm C}_{\rm beam}$ &
 ${\rm C}_{\rm off.}$ &  ${\rm C}_{\rm hel.}$ \\ \hline \hline
$m_t/2$ & $-0.364$   & 0.774    & 0.779  &   0.278
\\
$m_t$   & $-0.389$   & 0.806  & 0.813 &  0.311
\\
$2m_t$ & $-0.407$  & 0.829 & 0.836  &  0.331
\\ \hline
\end{tabular}
\end{center}
\caption{\it Dependence of the correlation coefficients,
computed with the PDF  of \protect\cite{CTEQ5},  on
$\mu=\mu_R=\mu_F$ at NLO.}\label{tab:mudep}
\end{table} 
\par
In Table \ref{tab:pdf}
we compare results for C using
different sets of PDF.
In the case of $p\bar{p}$ collisions at $\sqrt{s}=2$ TeV,
the spread of the results is larger
than the scale uncertainty given in Table \ref{tab:mudep}.
To a considerable extent this is due to an interesting feature of C,
namely the $q\bar{q}$ and $gg$ initial states
contribute to C with opposite signs.
Therefore the spin correlations are quite sensitive
to the relative weights of $q\bar{q}$ and $gg$ initiated $t\bar t$ events.
These weights depend in particular on the chosen set of PDF.
For example, one finds the following individual NLO contributions
for the helicity, beam, and off-diagonal correlation 
at the upgraded Tevatron: for the GRV98 (MRST98) PDF 
C$^{q\bar q}_{\rm hel.}=-0.443\ (-0.486)$, 
C$^{gg}_{\rm hel.}=+0.124\  (+0.075)$, 
C$^{q\bar q}_{\rm beam}=+0.802\ (+0.879)$, 
C$^{gg}_{\rm beam}=-0.068\  (-0.042)$, and
C$^{q\bar q}_{\rm off.}=+0.810\ (+0.889)$, 
C$^{gg}_{\rm off.}=-0.073\  (-0.044)$. This suggests that
accurate measurements  of the dilepton
distribution (\ref{eq:ddist1}), using different spin bases,
at the upgraded Tevatron may provide additional constraints
in the continuing
effort to improve 
the knowledge of the PDF.
\begin{table}[htb]
\begin{center}\renewcommand{\arraystretch}{1.5}
\begin{tabular}{|c|ccc|c|}\hline
& \multicolumn{3}{c|}{$p\bar p$ at $\sqrt{s}=2$~TeV}
& $p p$ at $\sqrt{s}=14$~TeV \\
\cline{2-5}
 PDF   & ${\rm C}_{\rm hel.}$ &  ${\rm C}_{\rm beam}$ &
 ${\rm C}_{\rm off.}$ &  ${\rm C}_{\rm hel.}$ \\ \hline \hline
GRV98 & $-0.325$   & 0.734    & 0.739  &  0.332
\\
CTEQ5   & $-0.389$   & 0.806  & 0.813 &  0.311
\\
MRST98 & $-0.417$  & 0.838 & 0.846  &  0.315
\\ \hline
\end{tabular}
\end{center}
\caption{\it Correlation coefficients ${\rm C}_{\rm hel.}$,
${\rm C}_{\rm beam}$, and ${\rm C}_{\rm off.}$
at NLO for $\mu_R=\mu_F=m_t$ and different sets
of parton distribution functions:  GRV98 \protect\cite{Gluck:1998xa}, 
CTEQ5 \protect\cite{CTEQ5}, and MRST98 (c-g)
\protect\cite{MRST98}.}\label{tab:pdf}
\end{table}
\par
Finally we have studied  the dependence of the C coefficients 
on the top quark mass. For this we have used again the CTEQ5 PDF
and set $\mu=m_t$. In the case of $p\bar{p}$ collisions
at $\sqrt{s}=2$ TeV, a variation of $m_t$ from 170 to 180 GeV
changes  ${\rm C}_{\rm hel.}$ from $-0.378$ to $-0.397$, 
${\rm C}_{\rm beam}$ from 0.790 to 0.817, and ${\rm C}_{\rm off.}$
from 0.797 to 0.822. At LHC energies, 
${\rm C}_{\rm hel.}$ changes by less than a percent. 
\par
The extension of our results to the ``lepton+jets'' and ``all jets'' decay
channels \cite{bbsu3} is straightforward. 
The ``lepton+jets'' channels should be particularly useful
for detecting $t\bar t$ spin correlations: although one looses top-spin
analyzing power one gains in statistics and the experimental 
reconstruction of the $t$ and $\bar t$ rest frames may also be facilitated.
\par
In conclusion we have analyzed, at next-to-leading order
in $\alpha_s$, the
hadronic production of $t\bar t$ quarks in a general spin configuration
and have
computed the dileptonic angular correlation coefficients  C that reflect
the degree of correlation between the $t$ and $\bar t$  spins.
Our  results for the Tevatron
show that the scale and in particular the
PDF uncertainties in the prediction
of the dileptonic angular
distribution  must be reduced before $t\bar t$ spin
correlations can be
used in a meaningful way to search for
relatively small effects
of new interactions that are, for example, 
not distinguished by violating parity
or CP invariance.  
Our results may also be useful to
learn more about the parton distributions in the proton at high energies.
For $pp$ collisions at $\sqrt{s}$ = 14 TeV the theoretical uncertainties
in the prediction of this distribution are
smaller and one may adopt the optimistic view that at the time the
LHC will be turned on further theoretical progress will have turned
top quark spin correlations  
into a precision tool for the analysis of $t\bar t$ events.
\\[1em]
{\bf Acknowledgments}
\\
We would like to thank S. Dittmaier, S. Moch  and  B. Pl\"umper
for useful discussions.

\end{document}